\documentclass[%
 reprint,
 superscriptaddress,
preprintnumbers,
nofootinbib,
 amsmath,amssymb,
 aps,
prl,
]{revtex4-1}

\usepackage{amsmath}
\usepackage{amssymb}
\usepackage{amsthm}
\usepackage{MnSymbol}
\usepackage{nicefrac}
\usepackage{amsfonts}
\usepackage{dsfont}

\usepackage[markup=nocolor]{changes}
\usepackage{natbib}

\usepackage{graphicx}

\usepackage[normalem]{ulem}

\usepackage{dcolumn}
\usepackage{bm}

\usepackage{color}
\usepackage[usenames,dvipsnames,svgnames,table]{xcolor}
\usepackage{slashed}
\usepackage{empheq}
\newcommand{\bea}{\begin{eqnarray}}
\newcommand{\eea}{\end{eqnarray}}
\newcommand{\be}{\begin{equation}}
\newcommand{\ee}{\end{equation}}

\begin{document}

\title{
Top mass from asymptotic safety
}
 
 \author{Astrid Eichhorn}
   \email{a.eichhorn@thphys.uni-heidelberg.de}
\affiliation{Institut f\"ur Theoretische
  Physik, Universit\"at Heidelberg, Philosophenweg 16, 69120
  Heidelberg, Germany}
\author{Aaron Held}
\email{a.held@thphys.uni-heidelberg.de}
\affiliation{Institut f\"ur Theoretische
  Physik, Universit\"at Heidelberg, Philosophenweg 16, 69120
  Heidelberg, Germany}

\begin{abstract}
We discover that asymptotically safe quantum gravity could predict the top-quark mass. For a  broad range of microscopic gravitational couplings, quantum gravity could provide an ultraviolet completion for the Standard Model by triggering asymptotic freedom in the gauge couplings and bottom Yukawa and asymptotic safety in the top-Yukawa and Higgs-quartic coupling. We find that in a part of this range, a difference of the top and bottom mass of approximately $170\, \rm GeV$ is generated and the Higgs mass is determined in terms of the top mass. Assuming no new physics below the Planck scale, we construct explicit Renormalization Group trajectories for Standard Model and gravitational couplings which link the transplanckian regime to the electroweak scale and yield a top pole mass of $M_\text{t,pole} \approx 171\, \rm GeV$. 
\end{abstract}

\pacs{Valid PACS appear here}

\maketitle

\begin{figure}[!t]
\includegraphics[width=\linewidth, clip=true, trim=10cm 5.8cm 12cm 5.6cm]{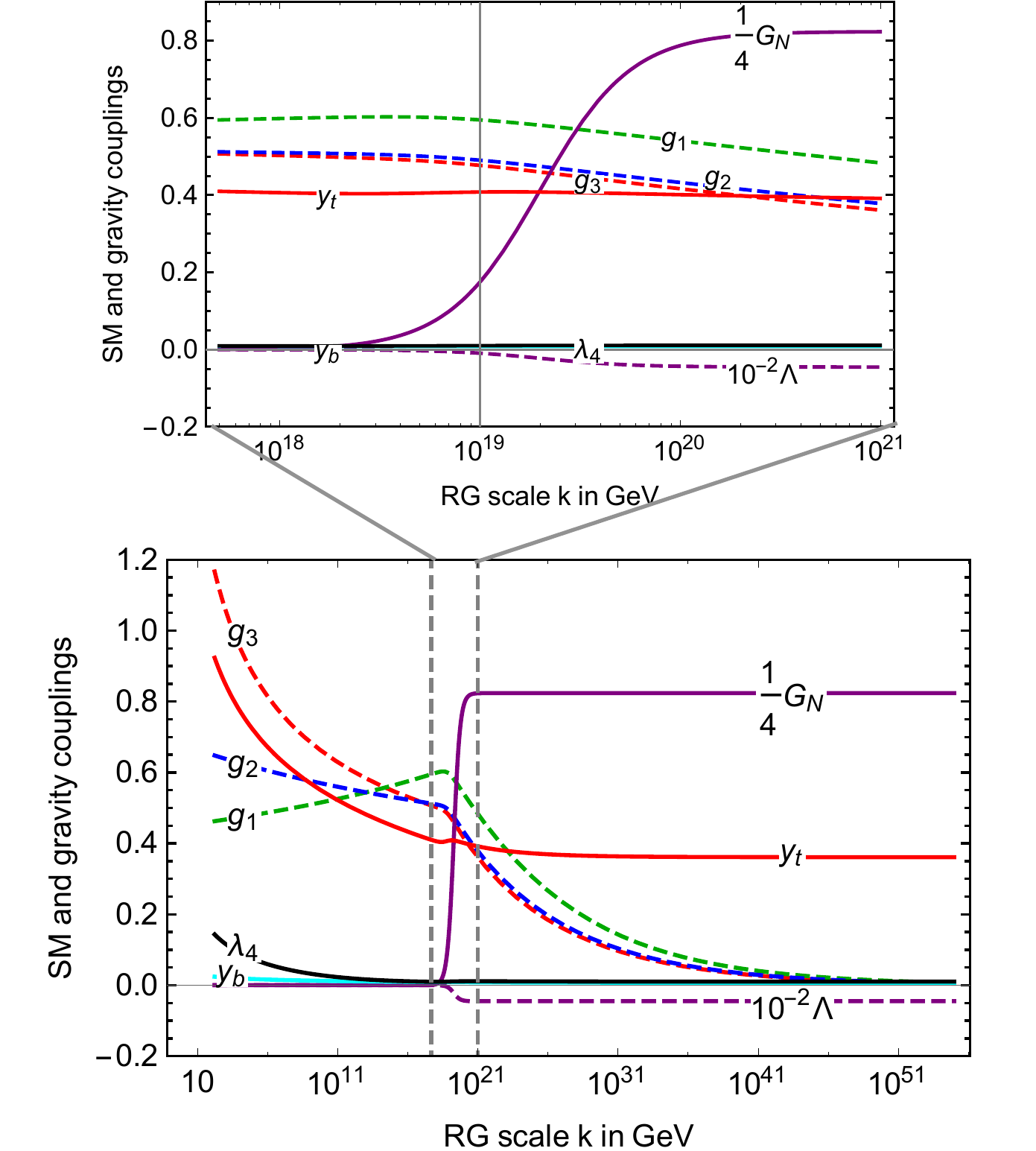}
\caption{\label{fig:flows} Running SM and gravity couplings in a UV-complete model, and a zoom into the regime close to the Planck scale, where the system smoothly transitions from pure SM running to the asymptotically safe running including gravitational fluctuations.}
\end{figure}

\noindent\emph{Open problems of the Standard Model}.
The Standard Model (SM) is a highly successful effective quantum field theory, viable up to the Planck scale
\cite{Bezrukov:2012sa,Buttazzo:2013uya}. 
Beyond,
it is expected to break down, due to
the triviality problem signaled by Landau poles in the Abelian hypercharge \cite{GellMann:1954fq} and the Higgs-Yukawa sector \cite{triviality}. Moreover, it contains 19 
free parameters. In particular, the 
various Yukawa couplings have to be set such that they generate the significant difference between the top-quark mass and the other fermion masses.\\
Asymptotically safe quantum fluctuations of gravity  \cite{Weinberg:1980gg,Reuter:1996cp}
might tame the ultraviolet (UV) divergences in the  Abelian gauge sector \cite{Harst:2011zx,Christiansen:2017gtg}. We discover that a quantum-gravity induced UV completion of the Higgs-Yukawa sector could at the same time eliminate two of the free parameters of the SM.  The resulting model would contain quantum gravity and all SM fields, be UV complete and have a higher predictive power than the SM, cf.~Fig.~\ref{fig:flows}: 
In a specific range
of microscopic gravitational couplings, the asymptotic safety scenario might 
predict the top mass in terms of the bottom mass, and generate a difference of $\sim 170\, \rm GeV$ between them.
\\
\noindent\emph{Asymptotic safety}
generalizes asymptotic freedom: The latter posits that a model evolves along a Renormalization Group (RG) trajectory which emanates from the free fixed point in the deep UV, whereas the former is based on a fixed point at finite values of the couplings. Both settings have in common that 
their free parameters are the relevant couplings which parameterize the deviation of the model from the fixed point. For instance, QCD with massless quarks only has one free parameter, namely the value of the gauge coupling at some energy scale. All other, so called irrelevant couplings arise as predictions of the model. They are forced to stay close to  their fixed-point values in the UV scaling regime and are determined by the relevant couplings at all scales. Conversely, any deviation  of an irrelevant coupling from this special value in the infrared (IR) results in a running that in simple approximations typically leads into a singularity, triggering a breakdown of the model. As an important example, the sizable value of the top Yukawa at the Planck scale within the SM cannot be reconciled with a free fixed point, as the coupling is marginally irrelevant and hence repelled by the free fixed point 
 when running further
into the UV. 
Our main result hinges on the fact that an asymptotically safe fixed point forces irrelevant couplings to stay at their \emph{non-zero} 
values in the UV scaling regime beyond the Planck scale. Thereby the scenario enforces a specific finite value of each irrelevant coupling at the Planck scale. Running from this unique 
value towards the IR
leads to a prediction of the low-energy value of the coupling. 
Within the present analysis, there is a non-trivial renormalization group fixed point which determines the top Yukawa and quartic Higgs couplings.

\noindent\emph{Quantum-gravity induced UV completion for the Standard Model}.
We focus on the Higgs-top-bottom sector of the SM coupled to the 
U(1) hypercharge with coupling $g_Y=\sqrt{3/5}g_1$, 
SU(2) with coupling $g_2$ and 
SU(3)-color gauge interactions with coupling $g_3$ under the impact of quantum gravity. In this work, we truncate the microscopic gravitational dynamics to an Einstein-Hilbert action parameterized by a dimensionless running Newton coupling $G_N = \bar{G}_N k^2$ and  dimensionless running cosmological constant $\Lambda = \bar{\Lambda}k^{-2}$. To evaluate the running with momentum scale $k$, we employ functional Renormalization Group tools \cite{Wetterich:1992yh}, see also \cite{Morris:1993qb,Berges:2000ew,Reuter:2012id} and \cite{Gies:2014xha} for a study of the Higgs-Yukawa sector. 
 There is strong evidence supporting the conjecture that the gravitational couplings $G_N$ and $\Lambda$ approach an asymptotically safe fixed point $(G_{N}^{\ast}, \Lambda^{\ast})$ at transplanckian momentum scales \cite{Reuter:1996cp,Litim:2003vp}, see \cite{Ambjorn:2012jv} for lattice studies.
The one-loop SM beta functions for the top-Yukawa and quartic Higgs 
coupling
including quantum fluctuations of gravity take the form
\bea
\beta_{y_t} &=&\frac{1}{32 \pi^2}\left(9 y_t^3 +3y_b^2 y_t +y_t\left(-
 16g_3^2 - \frac{9}{2}g_2^2-\frac{17}{10}g_1^2\right)\right)\nonumber \\[0.5ex]
&{}&+ G_N\, y_t\,f_y(\Lambda), \label{eq:betayt} \\[0.5ex]
\beta_{\lambda_4}&=& \frac{1}{8\pi^2} \Bigg(12 \lambda_4^2 +6 \lambda_4 \left(y_t^2 +y_b^2 - \frac{3}{20}g_1^2 -\frac{3}{4}g_2^2 \right) -3 \left(y_t^4
 +y_b^4\right)\nonumber\\[-0.5ex]
 &{}& \quad \quad
+\frac{27}{400}g_1^4 + \frac{9}{16}g_2^4  + \frac{9}{40}g_1^2 g_2^2 
 \Bigg)+ G_N\, \lambda_4\, f_{\lambda_4}(\Lambda),\label{eq:betalambda4}
\eea
where the quantum-gravity contributions 
in the above parameterization of metric fluctuations \cite{Eichhorn:2016esv,Eichhorn:2017eht,scalar_gravity} 
contain threshold effects
\bea
f_y(\Lambda)&=&\frac{96 +\Lambda (-235 + \Lambda (103+56\Lambda))}{12\pi(3+2\Lambda(-5+4\Lambda))^2},\\
f_{\lambda_4}(\Lambda)&=& \frac{165 - 8 \Lambda \left(61+\Lambda(-49+4\Lambda) \right)}{6\pi \left(3+2\Lambda(-5+4\Lambda) \right)^2}.
\eea
These are obtained in the Landau-gauge limit as in \cite{Eichhorn:2017eht} with an optimized cutoff function \cite{Litim:2001up}.
For the bottom-Yukawa,  the
corresponding beta function 
follows from Eq.~\eqref{eq:betayt} under the exchange $y_t \rightarrow y_b$ and  $17/10\, g_1 \rightarrow 1/2\, g_1$.
There are  indications
that asymptotically safe quantum gravity triggers asymptotic freedom in all gauge couplings, including the Abelian hypercharge coupling \cite{Daum:2009dn,Folkerts:2011jz,Harst:2011zx,Christiansen:2017gtg}.
Their one-loop beta functions with a gauge-group independent quantum-gravity contribution  in the above approximation and with the same choice of RG scheme and gauge parameters read
\bea
\beta_{g_1}&=&\frac{g_1^3}{16\pi^2}\frac{41}{10} 
 - G_N\, g_1f_g(\lambda)\;,  \label{eq:betag1}\\
\beta_{g_2}&=&- \frac{g_2^3}{16\pi^2}\frac{19}{6} 
- G_N\, g_2 f_g(\Lambda)\;,\nonumber\\
\beta_{g_3}&=&- \frac{g_3^3}{16\pi^2}7
 - G_N\, g_3 f_g(\Lambda), \quad  f_{g}(\Lambda)= \frac{5(1-4\Lambda)}{18 \pi(1-2\Lambda)^2}.\nonumber
\eea
According to these results, in the transplanckian regime, the logarithmic running of the gauge couplings is substituted by a power-law running towards asymptotic freedom, cf.~Fig.~\ref{fig:flows}, with a small exponent determined by the gravity couplings. The one-loop coefficients in Eq.~\eqref{eq:betag1} contain the effect of all SM matter fields.
\\
We find two distinct parts of the truncated gravitational parameter space: If the microscopic value of the cosmological constant -- which is not restricted by observations -- falls into the region $\Lambda^{\ast} \gtrsim-3.3$, then the top and bottom Yukawa only feature a free fixed point at which they are irrelevant. Thus, they remain stuck at zero in the transplanckian regime down to $M_{\rm Planck}$, implying a vanishing top and bottom mass at the electroweak scale.
Hence, we tentatively conclude that compatibility of asymptotically safe quantum gravity with the SM appears to require the microscopic value of the cosmological constant to fall into the other region \cite{Eichhorn:2016esv,Eichhorn:2017eht}. In enlarged truncations this regime might also be reached due to higher-order gravitational couplings \cite{Eichhorn:2017eht,Hamada:2017rvn}.
Within our approximation, the observationally viable regime lies at $f_y(\Lambda)<0$, where $y_t$ and $y_b$ become relevant at the free fixed point. Thus, non-zero IR values for those couplings become compatible with the latter.
Simultaneously, gravity induces an interacting fixed point at:
\bea
y_{t}^{\ast} &=&
\frac{\sqrt{32}\pi}{3}\sqrt{ -G_N\,f_y(\Lambda)} 
\;, \quad 
y_{b}^{\ast}= 0\;,  \quad 
g_{i}^{\ast} = 0\; \label{eq:FPyt},
\eea
realizing the scenario in \cite{Eichhorn:2017eht}. A finite fixed-point value for $\lambda_{4}$ follows  
by inserting $y_t^{\ast}$ from Eq.~\eqref{eq:FPyt} into the beta function in Eq.~\eqref{eq:betalambda4}.  This
equation has two solutions. We focus on the fixed point which features a higher predictive power due to the  irrelevance of the  quartic Higgs coupling, as proposed in \cite{Shaposhnikov:2009pv,Bezrukov:2012sa}. In our simple truncation, it lies at $\lambda_4^{\ast}>0$, which guarantees stability of the potential. The positive value of $\lambda_4$ at and beyond the Planck scale leads to an overestimation of the Higgs mass in the IR, as the experimentally determined value of the Higgs mass appears to lie slightly below the stability bound, see \cite{Bezrukov:2012sa,Degrassi:2012ry,Buttazzo:2013uya} and references therein. 
We hypothesize that effects beyond our truncation might shift that fixed point to
lower values of $\lambda_4$ while simultaneously guaranteeing global stability of the potential through the presence of higher-order interactions.\\
In our approximation all other Yukawas feature a free fixed point under the impact of quantum gravity, and can thus be chosen to match the experimentally determined small values.
As an adequate quantitative approximation, we set all but $y_b, y_t$ to zero at all scales.
\\
Studies of asymptotically safe gravity-matter systems have not yet reached a stage of quantitative convergence regarding the fixed-point values for the gravitational couplings  \cite{Dona:2013qba,Meibohm:2015twa,Dona:2015tnf,Eichhorn:2016vvy,Biemans:2017zca}. To reach the observationally viable regime in our parameterization $\Lambda^{\ast}\lesssim-3.3$ has to hold.  It is a critical question for the future, whether extended approximations that treat quantum fluctuations around the physical background will converge into this regime.
As highlighted in \cite{Eichhorn:2017eht}, truncations in pure gravity tend to yield fixed points  with
$\Lambda^{\ast} \geq 0$. In the observationally favored regime, off-shell fluctuations of the transverse-traceless mode no longer dominate the RG flow, as they typically do in pure-gravity truncations.
 \begin{figure}[t]
\includegraphics[width=\linewidth]{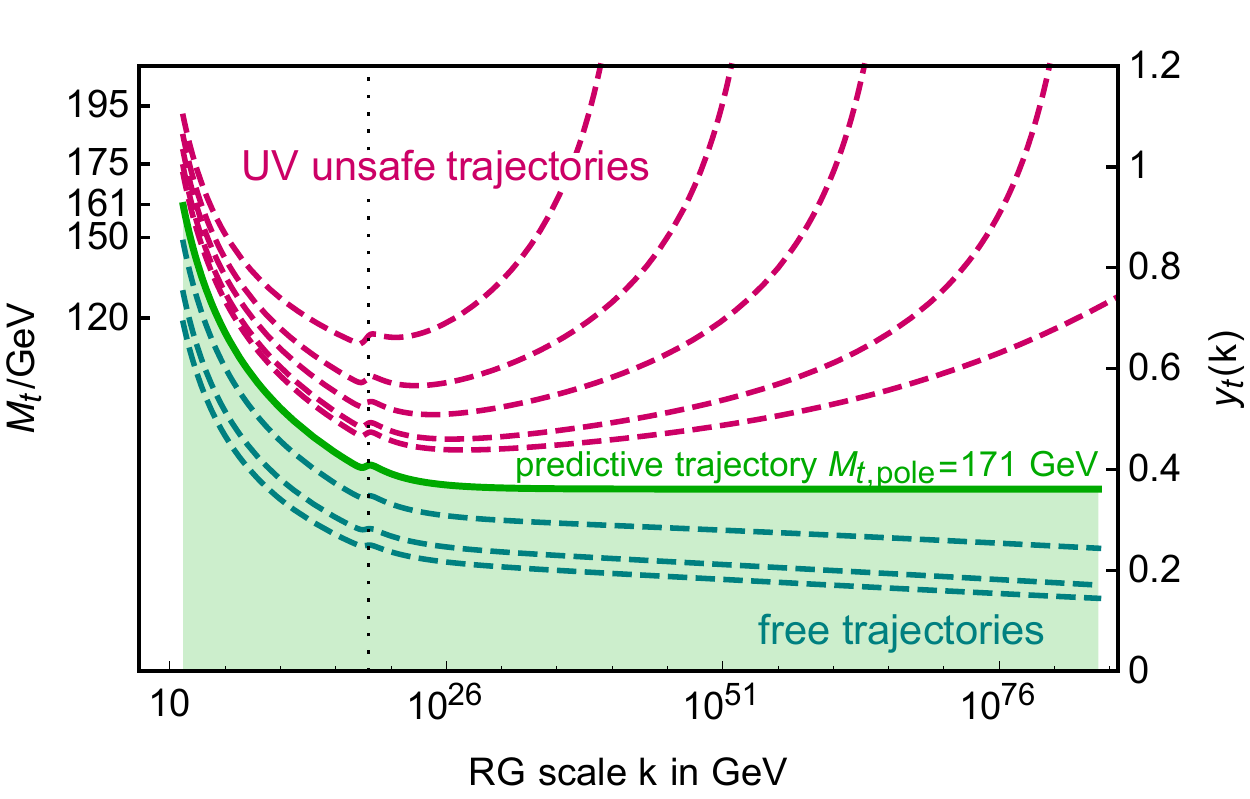}
\vspace*{-15pt}
\caption{\label{fig:upperbounds} Upper bound on the running top mass: A larger top mass leads to UV unsafe trajectories. From the upper bound, an asymptotically safe UV regime is reached. Below the upper bound, the Yukawa coupling becomes asymptotically free.}
\end{figure}
It is intriguing that a calculation of the microscopic values of the gravitational couplings under the impact of minimally coupled SM fields falls right into the preferred region \cite{Dona:2013qba}. For our quantitative analysis, we will employ the gravitational beta functions from \cite{Eichhorn:2016vvy} including the matter contributions from \cite{Dona:2013qba}, which provides beta functions in the Landau-gauge limit, reading
\bea
\beta_{G_N}&=& 2 \,G_N - G_N^2 \, f_{G_N}(\Lambda),\\[1ex] \nonumber
\beta_{\Lambda}&=& -2 \,\Lambda - \frac{G_N}{2\pi}\Bigl(7 - \frac{3}{2(3-4\Lambda)}+ N_W -\frac{N_S}{2} - N_V\\[-0.5ex]
&{}& -\frac{5}{ 2(1-2\Lambda)}- 
8\ln(3/2)
\Bigr) - G_N\,\Lambda \,f_{G_N}(\Lambda), \nonumber
\eea
with $N_S$ scalars, $N_V$ gauge fields and $N_W$ Weyl fermions ($N_S=4, N_V=12, N_W =45$ for the SM), and
\bea
f_{G_N}(\Lambda)&=&\frac{5}{6\pi(1-2\Lambda)}+ \frac{5}{3\pi(1-2\Lambda)^2} - \frac{1}{2\pi(3-4\Lambda)} \nonumber\\
&{}&+ \frac{{11}+32\ln(3/2)}{12\pi} 
 -\frac{N_W + N_S - 4N_V}{6\pi}. 
\eea
These provide asymptotically safe fixed-point values at
\bea 
G_{N}^{\ast}&=& 3.29,\; \Lambda^{\ast}=-4.51\,.\; \label{eq:fpvaluesgrav}
\eea
\noindent\emph{Upper bound on the top mass from asymptotic safety}.
Asymptotically safe quantum gravity with microscopic values according to Eq.~\eqref{eq:fpvaluesgrav} enforces upper bounds on the top- and bottom-quark mass: These arise due to the basin of attraction of the free fixed point at $y_{t}^{\ast} = 0 =y_{b}^{\ast}$. The basin is limited by interacting fixed points, cf.~Eq.~\eqref{eq:FPyt}. The resulting bounds  on the masses are more intuitive when considering the RG flow starting in the IR: For quark masses below the upper bound, the flow is attracted into the free fixed point, cf.~Fig.~\ref{fig:upperbounds}. Exactly at the upper bound, the couplings reach the interacting fixed point. In all cases, the fixed point for $\lambda_4$  -- which remains irrelevant -- can only be reached from one particular value of the Higgs mass, which is fixed in terms of the top mass.  For the bottom Yukawa, a further fixed point beyond that displayed in Eq.~\eqref{eq:FPyt}, at which $y_{b}^{\ast} >0$ enforces a similar upper bound. For masses beyond the upper bound, the Planck-scale values of the Yukawas are too large to approach any of the fixed points, cf.~Fig.~\ref{fig:upperbounds}, and the model breaks down at high scales, requiring new physics. 
\\
The free and interacting fixed points for the top Yukawa 
underlie two different possible UV completions for the SM plus gravity, with the latter having a higher predictive power.
For the remainder of this letter, we will focus on the asymptotically safe fixed point in Eq.~\eqref{eq:FPyt}, which results in a uniquely determined value for the top mass, as the top Yukawa is no longer a free parameter of the model.
\\

\noindent\emph{Top-bottom mass difference}.
At the fixed point in Eq.~\eqref{eq:FPyt} the relevant direction is not exactly aligned with the bottom Yukawa coupling. Instead the combination
\vspace*{-2pt}
\be
\vspace*{-1pt}
\tilde{y} = y_b - \frac{1}{5}y_t,\label{eq:reldir}
\ee
is relevant.
 Thus, any value of the bottom mass below the upper bound can be reached along a UV complete RG trajectory.
We choose it such that it matches the observed value of $M_b=4.18\, \rm GeV$. According to Eq.~\eqref{eq:reldir}, changing the bottom mass results in a slight change of the top mass. Given the gravitational fixed-point values in Eq.~\ref{eq:fpvaluesgrav}, realistic values for the bottom mass enforce a significant mass difference to the top quark. It
is automatically of a similar size to that of the SM. Specifically, $M_{t,\,\rm pole }-M_b \geq 160\, \rm GeV$ holds for $M_b \lesssim 11\, \rm GeV$.\\
Alternatively, a mass difference could be generated from the finite fixed-point value for the Abelian gauge coupling that arises in Eq.~\eqref{eq:betag1}, as found in \cite{Harst:2011zx,Eichhorn:2017lry}, yielding an additional interacting fixed point with distinct values for the top and bottom Yukawa \cite{Held20175}.\\
We stress that the electroweak scale is an input of our calculation: The Higgs mass-parameter remains relevant, and thus the vacuum expectation value of $v = 246\, \rm GeV$ can be reached by choosing an appropriate trajectory. For the fermions, a separation of their mass scale from the Planck scale is possible as fermion masses are protected by chiral symmetry and remain so under the impact of asymptotically safe quantum gravity \cite{Eichhorn:2011pc,Eichhorn:2016vvy}. Hence, their masses are generated by electroweak symmetry breaking, and the precise values are determined by the Yukawa couplings.
\\

\noindent\emph{Renormalization Group flow of the Standard Model and quantum gravity.}
Starting from the asymptotically safe regime in Eq.~\eqref{eq:FPyt} which, given  
values in Eq.~\eqref{eq:fpvaluesgrav}  lies at
\bea
 y_{b}^{\ast}&=&0,\; y_{t}^{\ast}= 0.36,\; \lambda_{4}^{\ast}=0.01,\; g_{i}^{\ast}=0,\label{eq:fpvalues}
 \eea
we obtain $M_t = 161\, \rm GeV$ for the running top mass along the predictive trajectory, cf.~Fig.~\ref{fig:flows}. Adding the correction between the running mass and the pole mass \cite{Hempfling:1994ar,Buttazzo:2013uya}, we obtain an estimate of the top pole mass of $M_\text{t,\,pole} =171$, falling in the vicinity of the experimentally determined value \cite{Khachatryan:2015hba}. \\
Above the Planck scale, the Yukawa and  quartic Higgs already exhibit a slow running away from their fixed-point values. This is a consequence of the presence of the gauge couplings which pull the 
quartic Higgs and the Yukawa 
along as soon as they deviate from their free values in the deep UV.
The bump in the vicinity of the Planck scale is a consequence of a rapid transition from the semi-classical gravity regime to the fixed-point regime, cf.~Fig.~\ref{fig:flows}. Below the Planck scale, an IR attractive fixed point in the ratio of top Yukawa and gauge coupling and the ratio of the Higgs quartic coupling and  top Yukawa \cite{Wetterich:1981ir,Wetterich:1987az,Bornholdt:1992up} act as an attractor of the flow, yielding a Higgs mass of $M_h= 132\, \rm GeV$ in our simple truncation. Higher-order interactions 
might reconcile global stability of the potential with a Higgs mass agreeing with the experimental result \cite{Aad:2015zhl}.
\begin{figure}[t]
\vspace*{4pt}
\begin{flushleft}
\includegraphics[width=0.9\linewidth]{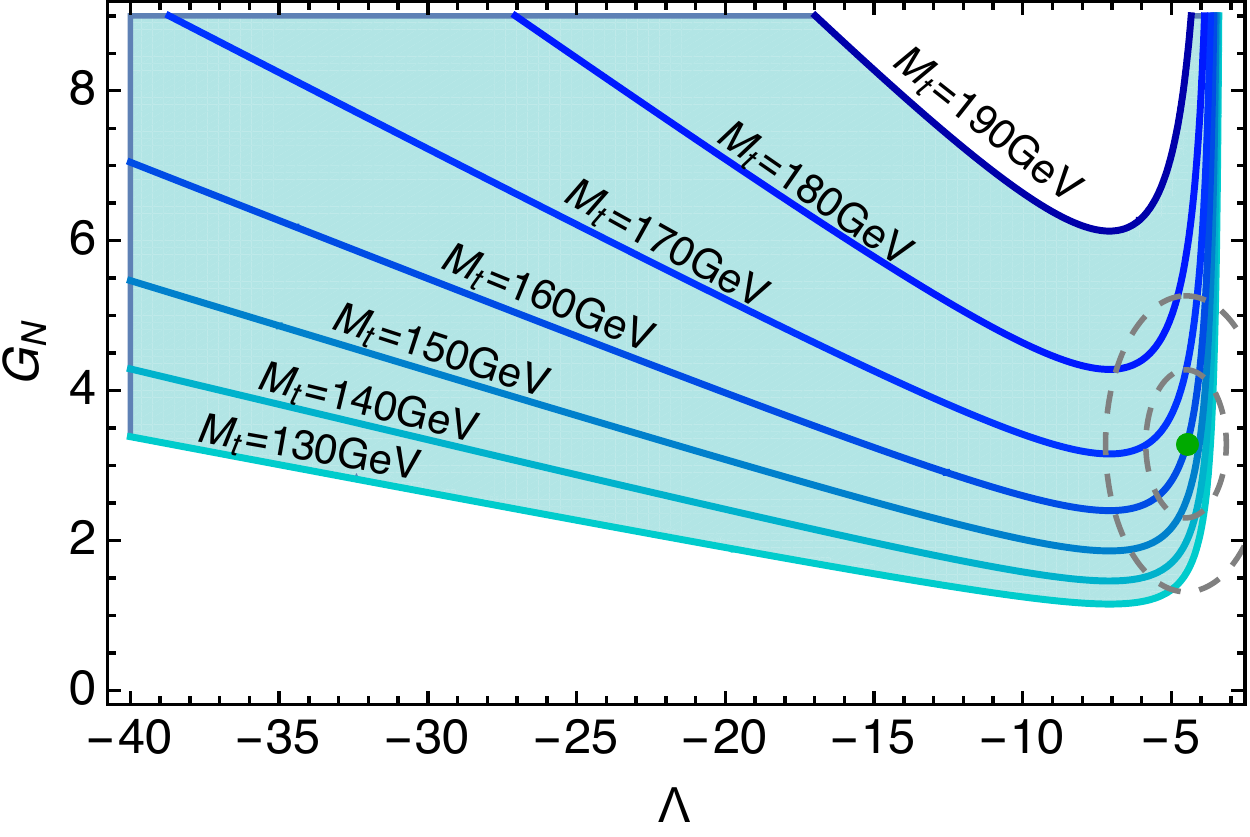}
\end{flushleft}
\vspace*{-20pt}
\caption{\label{fig:glambda}Running top mass as a function of the microscopic values of the gravity couplings. The green dot is the result from Eq.~\eqref{eq:fpvaluesgrav} and the gray dashed curves indicate an estimate for a systematic error within the approximation.}
\end{figure}
\\
\noindent The predicted values of the masses depend on the microscopic values of the gravity couplings, which are only known in approximations. Further studies are required to determine whether those values will converge in the regime where a top mass prediction is possible. For the following, we assume that this will be the case. Varying the gravitational couplings away from the values in Eq.~\eqref{eq:fpvalues}, cf.~Fig.~\ref{fig:glambda}, 
we obtain a range of masses. We consider only values for which the scalar potential in our truncation is stable and for which $M_h<M_t$, yielding the light blue region in Fig.~\ref{fig:glambda}. 
We estimate a systematic error within our approximation under the assumption that extended truncations will converge in a similar regime. Including variations of $G_N^{\ast}, \Lambda^{\ast}$ induced by changes of the regulator underlying the functional RG implementation according to Tab.~I in \cite{Dona:2013qba} leads to changes of up to $60\%$ in the fixed-point values. In Fig.~\ref{fig:glambda} we include an ellipse with $60\%$ overall spread, resulting in a rather significant variation for $M_t$. 
The difference between the running and the pole mass, that receives QCD corrections \cite{Hempfling:1994ar,Buttazzo:2013uya} is not included in the values shown in Fig.~\ref{fig:glambda}. As a more conservative error estimate, the larger ellipse with a radial deviation of $60\%$ strongly reinforces the need for extended truncations.
\\
\noindent 
Through their quantum fluctuations, right-handed neutrinos (and the axion) impact the microscopic values of the gravitational couplings \cite{Dona:2013qba}. Accounting for this shift, we obtain a running top mass of $ M_t = 182\, ( 185
)\, \rm GeV$. It is intriguing that in our setting, the values of the top and the Higgs mass are actually sensitive to new degrees of freedom at arbitrarily high scales which need only be coupled gravitationally. Within asymptotic safety it might accordingly be possible to draw conclusions about new physics at \emph{arbitrary} scales, once the systematic error induced by truncations is under control. This assumes that the gravity fixed-point values converge to the regime that we investigate at $\Lambda^{\ast}\lesssim-3.3$.
\\

\noindent\emph {Conclusions:}
We explore a regime of asymptotically safe quantum gravity, in which a UV completion for the SM is triggered. The top mass and Higgs mass arise as predictions in this setting. In particular, fixing the bottom mass to its observed value enforces a mass difference between top and bottom of approximately $ 170\, \rm GeV$.
\\
In more detail, the following holds in a truncation of the RG flow:
All gauge couplings of the SM become asymptotically free under the impact of quantum gravity fluctuations \cite{Daum:2009dn,Folkerts:2011jz,Harst:2011zx,Christiansen:2017gtg}. This includes the Abelian gauge coupling, which 
exhibits a power-law running towards asymptotic freedom in the transplanckian regime.
\\
We discover an upper bound on the top mass: It is set by a quantum-gravity induced interacting fixed point for the top Yukawa. For values below the upper bound the top Yukawa exhibits a power-law running towards asymptotic freedom. Exactly at the upper bound, the UV regime is asymptotically safe. Beyond the upper bound, the top Yukawa diverges along UV unsafe trajectories.
\\
We then focus on the asymptotically safe fixed point for the top Yukawa. On the trajectory that emanates from the asymptotically safe regime in Eq.~\eqref{eq:FPyt} the top mass is predicted in terms of the bottom mass. Choosing $M_b=4.18\, \rm GeV$ automatically enforces a mass difference with the top, that is quantitatively close to the observed size of $M_{t,\, \rm pole}-M_b\sim 170\, \rm GeV$, and specifically yields a top pole mass $M_{t,\, \rm pole} \approx 171\, \rm GeV$. As proposed in \cite{Shaposhnikov:2009pv,Bezrukov:2012sa}, the Higgs mass also becomes a prediction.\\
We stress that our results arise in a truncation of the RG flow that is limited to  the surmised leading-order effects of quantum gravity on matter. 
In particular, convergence for the microscopic values of the gravity couplings under the inclusion of matter effects has not yet been achieved, and might require significant extensions of the truncation. We caution that the full system might feature fixed-point values in the regime $\Lambda^{\ast}>-3.3$.
Quantitatively precise results require extensions of the truncation in the gravity sector, and a matching at the Planck scale to the NNLO running of the SM. 
The uncovered predictive UV completion of the SM should give a strong incentive to advance in this direction.
\\
\noindent\emph{Acknowledgements}
We acknowledge insightful discussions with H.~Gies, J.~M.~Pawlowski and C.~Wetterich and would also like to thank H.~Gies and J.~M.~Pawlowski for valuable comments on the manuscript. We thank M.~Reichert for making us aware of an error in one of the beta functions in a previous version.
This research is supported by an Emmy-Noether fellowship through the DFG under grant no.~EI-1037-1.
A.~Held is also supported by a scholarship of the Studienstiftung des deutschen Volkes.

\end{document}